\documentclass[twocolumn]{aastex63}

\submitjournal{ApJ Letters}

\shorttitle{KELT-9b Phase Curve}
\shortauthors{Mansfield et al.}

\begin{document}

\title{Evidence for H$_{2}$ Dissociation and Recombination Heat Transport in the Atmosphere of KELT-9b} %KELT-9b: that girl is on fire

\correspondingauthor{Megan Mansfield}
\email{meganmansfield@uchicago.edu}

\author{Megan Mansfield}
\affiliation{Department of Geophysical Sciences, University of Chicago, Chicago, IL 60637, USA}

\author{Jacob L. Bean}
\affiliation{Department of Astronomy \& Astrophysics, University of Chicago, Chicago, IL 60637, USA}

\author{Kevin B. Stevenson}
\affiliation{Space Telescope Science Institute, Baltimore, MD 21218, USA}

\author{Thaddeus D. Komacek}
\affiliation{Department of Geophysical Sciences, University of Chicago, Chicago, IL 60637, USA}

\author{Taylor J. Bell}
\affiliation{Department of Physics, McGill University, Montr\'{e}al, QC H3A 2T8, Canada}

\author{Xianyu Tan}
\affiliation{Atmospheric, Ocean, and Planetary Physics, Clarendon Laboratory, Department of Physics, University of Oxford, Oxford OX1 3PU, United Kingdom}

\author{Matej Malik}
\affiliation{Department of Astronomy, University of Maryland, College Park, MD 20742, USA}

\author{Thomas G. Beatty}
\affiliation{Department of Astronomy and Steward Observatory, University of Arizona, Tucson, AZ 85721}

\author{Ian Wong}
\affiliation{Department of Earth, Atmospheric, and Planetary Sciences, Massachusetts Institute of Technology, Cambridge, MA 02139}

\author{Nicolas B. Cowan}
\affiliation{Department of Physics, McGill University, Montr\'{e}al, QC H3A 2T8, Canada}
\affiliation{Department of Earth \& Planetary Sciences, McGill University, Montr\'{e}al, QC H3A 2T8, Canada}

\author{Lisa Dang}
\affiliation{Department of Physics, McGill University, Montr\'{e}al, QC H3A 2T8, Canada}

\author{Jean-Michel D\'{e}sert}
\affiliation{Anton Pannekoek Institute for Astronomy, University of Amsterdam, 1090 GE Amsterdam, Netherlands}

\author{Jonathan J. Fortney}
\affiliation{Department of Astronomy and Astrophysics, University of California, Santa Cruz, CA 95064, USA}

\author{B. Scott Gaudi}
\affiliation{Department of Astronomy, The Ohio State University, Columbus, OH 43210}

\author{Dylan Keating}
\affiliation{Department of Physics, McGill University, Montr\'{e}al, QC H3A 2T8, Canada}

\author{Eliza M.-R. Kempton}
\affiliation{Department of Astronomy, University of Maryland, College Park, MD 20742, USA}
\affiliation{Department of Physics, Grinnell College, Grinnell, IA 50112, USA}

\author{Laura Kreidberg}
\affiliation{Harvard-Smithsonian Center for Astrophysics, Harvard University, Cambridge, MA 02138, USA}

\author{Michael R. Line}
\affiliation{School of Earth and Space Exploration, Arizona State University, Tempe, AZ 85281, USA}

\author{Vivien Parmentier}
\affiliation{Atmospheric, Ocean, and Planetary Physics, Clarendon Laboratory, Department of Physics, University of Oxford, Oxford OX1 3PU, United Kingdom}

\author{Keivan G. Stassun}
\affiliation{Department of Physics and Astronomy, Vanderbilt University, Nashville, TN 37235}

\author{Mark R. Swain}
\affiliation{Jet Propulsion Laboratory, California Institute of Technology, Pasadena, CA 91109, USA}

\author{Robert T. Zellem}
\affiliation{Jet Propulsion Laboratory, California Institute of Technology, Pasadena, CA 91109, USA}

\begin{abstract}
Phase curve observations provide an opportunity to study the energy budgets of exoplanets by quantifying the amount of heat redistributed from their daysides to their nightsides. Theories of phase curves for hot Jupiters have focused on the balance between radiation and dynamics as the primary parameter controlling heat redistribution. However, recent phase curves have shown deviations from the trends that emerge from this theory, which has led to work on additional processes that may affect hot Jupiter energy budgets. One such process, molecular hydrogen dissociation and recombination, can enhance energy redistribution on ultra-hot Jupiters with temperatures above $\sim2000$~K. In order to study the impact of H$_{2}$ dissociation on ultra-hot Jupiters, we present a phase curve of KELT-9b observed with the \textit{Spitzer Space Telescope} at 4.5~$\mu$m. KELT-9b is the hottest known transiting planet, with a 4.5-$\mu$m dayside brightness temperature of $4566^{+140}_{-136}$~K and a nightside temperature of $2556^{+101}_{-97}$~K. We observe a phase curve amplitude of $0.609 \pm 0.020$ and an offset of $18.7^{+2.1}_{-2.3} \degr$. The observed amplitude is too small to be explained by a simple balance between radiation and advection. General circulation models (GCMs) and an energy balance model that include the effects of H$_{2}$ dissociation and recombination provide a better match to the data. The GCMs, however, predict a maximum phase offset of $5\degr$, which disagrees with our observations at $>5\sigma$ confidence. This discrepancy may be due to magnetic effects in the planet's highly ionized atmosphere.
\end{abstract}


\section{Introduction}

Hot Jupiter phase curve observations have led to a wealth of data on energy transport in highly-irradiated planets \citep{Parmentier2018a}. This information has spurred the development of theories to describe the resulting trends. The most influential hypothesis has been that the irradiation level is the primary factor controlling energy transport, with hotter planets having shorter radiative timescales and thus less heat redistribution \citep[e.g.,][]{Showman2002,Rauscher2010,Heng2011,Cowan2011}. Lower heat redistribution would lead to increasingly larger phase curve amplitudes and smaller offsets. These trends with irradiation temperature are robust predictions that are born out in models with varying levels of sophistication \citep[e.g.,][]{Komacek2016}.

Recent phase curve observations, however, have shown deviation from these trends, which suggests that the radiative timescale may not be the only important factor controlling heat redistribution on hot Jupiters \citep[e.g.,][]{Zhang2018,Keating2019,Arcangeli2019}. In particular, ultra-hot Jupiters with temperatures $\gtrsim2000$~K should have additional important physics because they are hot enough that H$_{2}$ dissociates into hydrogen atoms on their daysides and recombines near the terminator \citep{Bell2018,Komacek2018,Parmentier2018b}. This process is predicted to distribute significant energy in a manner similar to latent cooling from water evaporation, with heat deposited in the regions where H recombines into H$_{2}$ \citep{Bell2018}. Such heat redistribution should lead to smaller phase curve amplitudes \citep{Bell2018,Komacek2018}. H$_{2}$ dissociation also provides a source of hydrogen atoms for the production of H$^{-}$, which is an important opacity source for ultra-hot Jupiters \citep{Arcangeli2018}.

In order to test predictions for energy transport in ultra-hot Jupiters, we present a phase curve of the transiting planet \object{KELT-9b} observed with the \textit{Spitzer Space Telescope} at 4.5~$\mu$m. KELT-9b is the hottest known planet, with a dayside temperature of $\sim4500$~K \citep{Gaudi2017}. This ultra-hot planet has been shown previously to contain neutral and ionized metals \citep{Hoeijmakers2018,Hoeijmakers2019}, and it is predicted to be heavily influenced by H$_{2}$ dissociation/recombination \citep{Bell2018,Komacek2018,Kitzmann2018,Lothringer:2018aa}.
It is also predicted to be too hot for clouds to form, even on the nightside, which simplifies potential interpretations of its phase curve \citep{Kitzmann2018}. We describe our observations and data reduction process in Section~\ref{sec:obs}. We compare our observations to a set of general circulation models (GCMs) in Section~\ref{sec:gcm} and energy balance models in Section~\ref{sec:ebm}. We discuss our results in Section~\ref{sec:discuss}.

\section{Observations and Data Reduction}
\label{sec:obs}

\begin{figure}
    \centering
    \includegraphics[width=\linewidth]{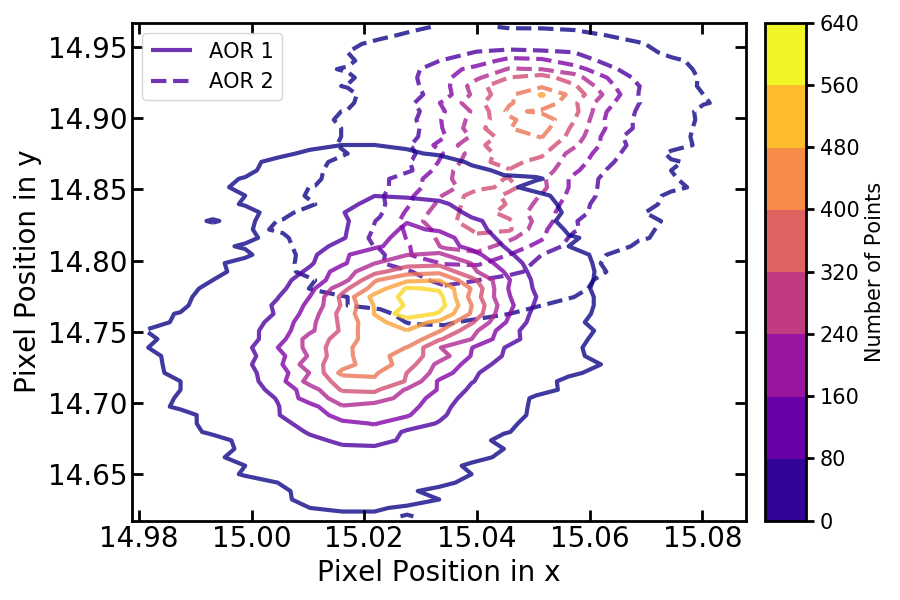}
    \caption{Contour plots showing the pointing in each of the two AORs. AOR 1 is in solid contours and AOR 2 is in dashed contours. The two pointings overlap significantly, allowing construction of an accurate pixel sensitivity map spanning the entire observation period.}
    \label{fig:pointing}
\end{figure}

We observed a single phase curve of KELT-9b with the InfraRed Array Camera (IRAC) at 4.5~$\mu$m on October 22-24, 2018 as part of a Cycle 14 large program (program ID: 14059). We used the subarray mode with 0.4-second frame times. Before beginning science observations, we performed a standard 30-minute pre-observation using the PCRS peak-up to mitigate spacecraft drift. Science observations were divided into two contiguous astronomical observation requests (AORs), which lasted for 22.3 and 18.6~hr, respectively. The two AORs had significant overlap in pointing, as shown in Figure \ref{fig:pointing}, and this observation had the most stable pointing overall of the nine phase curves observed to date in program~14059. A total of 371,392 frames were observed. We chose not to analyze the 30-minute pre-observation because it fell on a region of the detector that has little overlap with the two science AORs.

\begin{figure}
    \centering
   \includegraphics[width=0.7\linewidth]{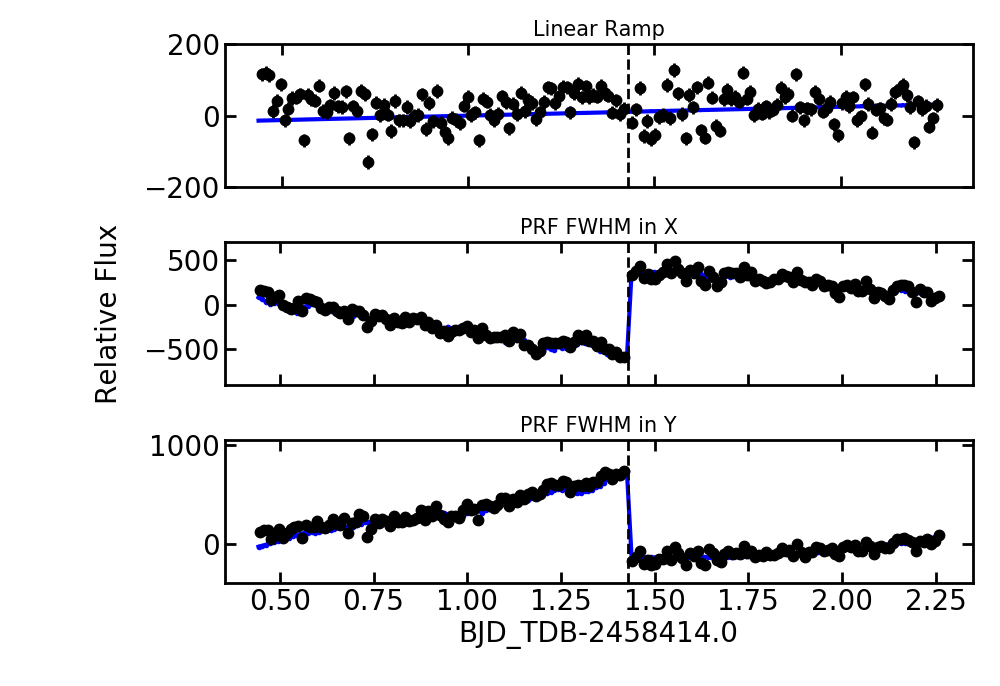}
    \caption{Trends over time of the main parameters we decorrelate against: a long-term linear ramp (top) and the PRF FWHM in the x (middle) and y (bottom) directions. Blue lines show the best-fit models, and black points show the data. The vertical dashed line indicates the break between the two AORs, where a slight change in telescope pointing causes a jump in the PRF FWHM.}
    \label{fig:decorr}
\end{figure}

\begin{figure}
    \centering
    \includegraphics[width=0.7\linewidth]{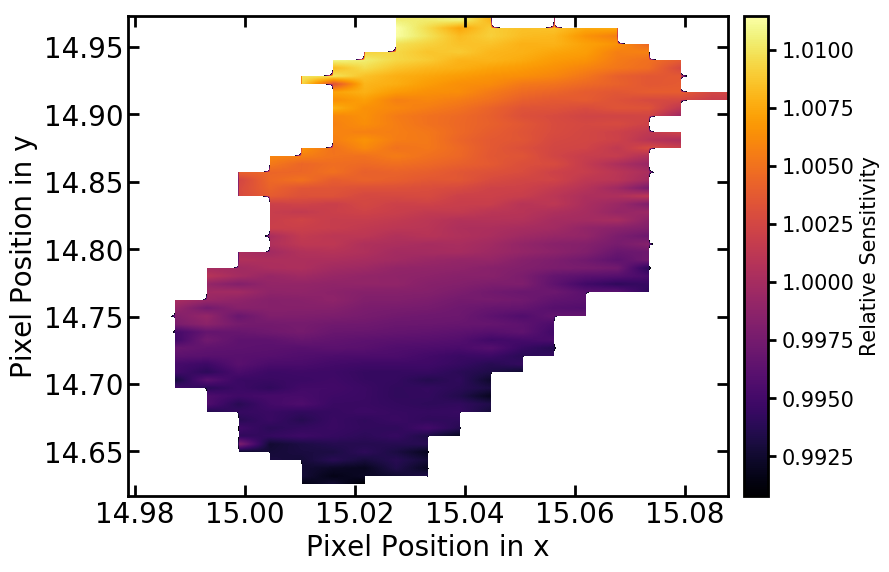}
    \caption{BLISS map illustrating the position-dependent pixel sensitivity. Yellow (purple) colors indicate regions of the pixel that are more (less) sensitive. Sensitivity drops with distance from the pixel center at (15.0,15.0), as expected.}
    \label{fig:BLISS}
\end{figure}

We reduced the data using the Photometry for Orbits, Eclipses, and Transits (POET) pipeline \citep{Campo2011,Stevenson2012,Cubillos2013}. We tested a range of fixed and variable aperture sizes \citep{Lanotte2014} and found the smallest scatter was achieved with a fixed circular aperture with a radius of 2.5 pixels. We binned sets of 4 images together for the data reduction because we found that this is the smallest bin size that produces a strong constraint on the Point Response Function Full Width at Half-Maximum (PRF FWHM). We modeled position-dependent systematics using Bilinearly Interpolated Subpixel Sensitivity (BLISS) mapping with a step size of 0.006 pixels \citep{Stevenson2012}. The BLISS map is shown in Figure~\ref{fig:BLISS}.
We also decorrelated against the PRF FWHM, as this has been recently shown to improve the fit quality \citep{Mendonca2018}. We tested models with linear, quadratic, and cubic dependences on the PRF FWHM in both the x and y directions, as well as a model without this dependency, and found that a linear model in both directions provides the preferred solution as determined by the Bayesian Information Criterion (BIC)\footnote{The usefulness of the BIC is limited in this case because BLISS mapping involves several free parameters that are not counted, but the $\Delta$BIC can still be used to differentiate between models with different numbers of parameters.
}. Additionally, we modeled a long-term linear trend over the entire phase curve. We tested a quadratic long-term trend and found that the linear trend is favored, with a $\Delta$BIC$=8$. Figure \ref{fig:decorr} shows the trends over time of the parameters we decorrelate against.

We modeled the phase-dependent emission of KELT-9b using a two-term sinusoid of the form
\begin{equation}
    F_{p}=A_{1}\cos\left[\frac{2\pi(t-t_{1})}{p}\right]+A_{2}\cos\left[\frac{4\pi(t-t_{2})}{p}\right],
    \label{eq:twocos}
\end{equation}
where $t$ is time, $p=1.4811$\,d is the orbital period, and $A_{1}$, $A_{2}$, $t_{1}$, and $t_{2}$ are free parameters. The second sinusoid allows a fit to an asymmetric phase curve and has been used to model several other phase curves \citep[e.g.,][]{Knutson2012,Stevenson2017}. We tested models using one or three sinusoids, but found a model with two sinusoids is preferred with a $\Delta$BIC$=8$ compared to a one-term model and a $\Delta$BIC$=21$ compared to a three-term model. This test gives us additional confidence in our analysis because third-order harmonics should not exist on a static map \citep{Cowan2017} We additionally tested for the presence of ellipsoidal variations by fixing the offset $t_{2}$ 
to a time chosen such that the sinusoid has maxima at quadrature and minima at transit and eclipse \citep{Shporer2017}. We found no evidence for ellipsoidal variations above the noise level of the observations, and so  
left $t_{2}$ as a free parameter in the final fit. We fit the transit and eclipses using the model of \citet{Mandel2002}, and used a linear model of stellar limb darkening during the transit.

\citet{Wong2019} found an additional periodicity in \textit{TESS} phase curves of KELT-9b with a period of $\approx7.6$~hr and semi-amplitude of 117\,ppm, which they attribute to stellar pulsations. We confirm the presence of this periodicity through a periodogram analysis of the residuals to our fit. We therefore include a model for this periodicity in our analysis using the equation
\begin{equation}
    F_{*}=1+\alpha \sin \left[\frac{2\pi(t-T_{0})}{\Pi}\right]+\beta \cos \left[\frac{2\pi(t-T_{0})}{\Pi}\right],
\end{equation}
where $T_{0}$ is the transit midpoint and $\alpha$, $\beta$, and $\Pi$ are free parameters. We find that including these pulsations has an almost negligible influence on our fitted phase curve parameters, which is not surprising because the planet's thermal emission at these infrared wavelengths is more than ten times larger than the stellar pulsation signal. Nevertheless, we retrieve a period and amplitude for the signal consistent with that of \citet{Wong2019}, and including it in our model removes some of the correlated noise present in the raw phase curve.

\begin{figure*}
    \centering
    \includegraphics[width=\linewidth]{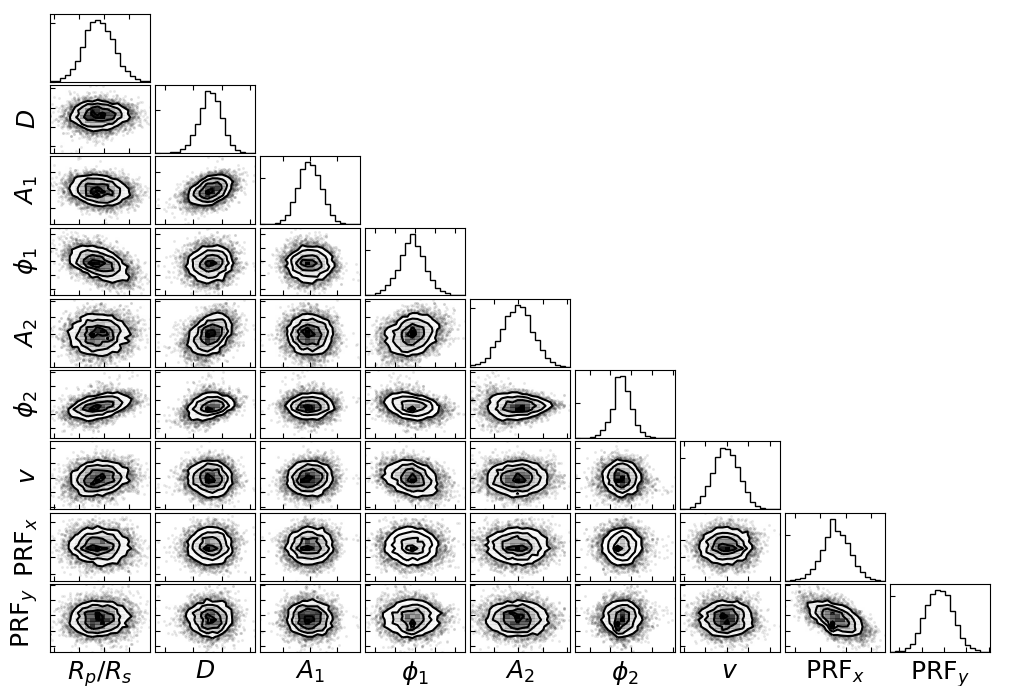}
    \caption{Pairs plot showing the key parameters we fit for. Parameter abbreviations are defined in Table~\ref{tab:params} and Equation~\ref{eq:twocos}.}
    \label{fig:pairs}
\end{figure*}

We estimated the parameters using a Differential Evolution Markov Chain Monte Carlo (MCMC) fit \citep{terBraak2008} with uniform priors for all parameters. Figure~\ref{fig:pairs} shows a pairs plot from the MCMC fit and 
Table~\ref{tab:params} lists the values of all fitted parameters. The data exhibit time-correlated noise, so we followed the red noise correction procedure of \citet{Diamond2014} and included this effect in our uncertainty estimates using the wavelet analysis described by \citet{Carter2009}. We initially fit for the $\gamma$ parameter described in \citet{Carter2009}, and then in the final MCMC fixed it to the best-fit value of $\gamma=0.58$.

\begin{figure}
    \centering
    \includegraphics[width=\linewidth]{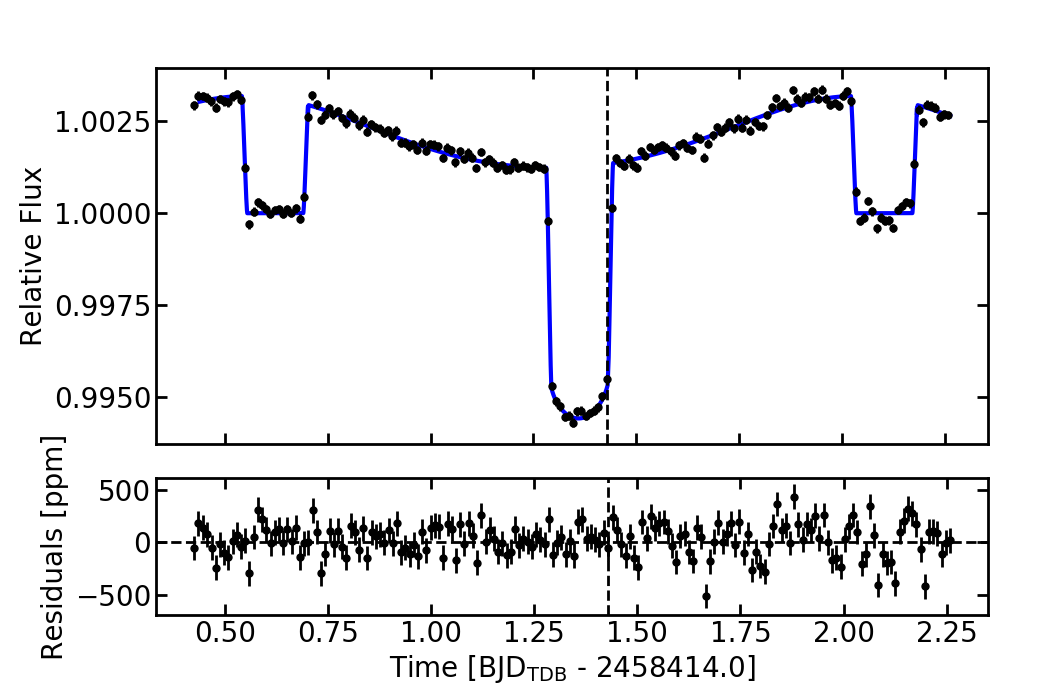}
    \caption{Phase curve of KELT-9b (black points) and best-fit model (blue line). The dashed vertical line shows the break between the two AORs. The lower panel shows residuals to the model fit. The data were binned into sets of 4 frames for the analysis, but for clarity the data here and in the other figures are shown binned into 180 bins of 2037 frames each ($\approx15$~min/bin). The RMS of the residuals at this binning is 118\,ppm, and the photon noise is 60\,ppm.}
    \label{fig:phase}
\end{figure}

The detrended phase curve is shown in Figure~\ref{fig:phase}. The RMS of the residuals when binning the data into 180 points ($\approx15$~min/bin) is 118\,ppm, and the photon noise is 60\,ppm. Table~\ref{tab:params} lists several parameters derived from the phase curve, including the dayside and nightside brightness temperatures ($4566^{+140}_{-136}$
\,K and $2556^{+101}_{-97}$
\,K, respectively), which were derived using PHOENIX models for the star \citep{Husser2013}. The error on our derived temperatures incorporates the relatively large error on the stellar effective temperature of KELT-9 \citep[$T_{\rm eff}=10170 \pm 450$~K,][]{Gaudi2017}. The dayside temperature we observe at 4.5~$\mu$m is consistent with the temperature of $4600\pm150$~K observed in the \textit{z}' band \citep{Gaudi2017}, which is expected from some 1D models of KELT-9b's atmosphere given the measurement uncertainties \citep{malik19}. We also derived a day-night temperature contrast of \begin{equation}
 A_{T}=\frac{T_\mathrm{day} - T_\mathrm{night}}{T_\mathrm{day}}   = 0.440^{+0.017}_{-0.016},
\end{equation}
an amplitude of 
\begin{equation}
    A=\frac{F_{max}-F_{min}}{F_{max}}=0.609 \pm 0.020,
\end{equation}
and a phase offset of $18.7^{+2.1}_{-2.3}\degr$.

To ensure the robustness of our results, we tested analyzing the two AORs separately and analyzing a phase curve with the bump in the data at $\mathrm{BJD}_{\mathrm{TDB}} \approx 2458415.8$~d masked out, and in all cases derived phase offsets and amplitudes that were consistent to within $1\sigma$. These data were also analyzed independently by T.~Beatty and D.~Keating to test for dependence on the data reduction method. The resulting amplitudes and phase offsets agreed within $1.5\sigma$. %Beatty values all within 1 sigma. Keating values offset within 1 sigma, amplitude within 1.5 sigma
A combined analysis of these data with a \textit{Spitzer} 3.6~$\mu$m phase curve of KELT-9b will be presented in a future paper (T.~Beatty et al. in prep.).

\begin{table}[]
    \centering
    \begin{tabular}{>{\centering\arraybackslash}p{5.4 cm} c}
    \hline \hline
        \textbf{Fitted Parameters} & \textbf{Value} \\
        \hline
        Transit Midpoint [BJD$_{\text{TDB}}$] & $2458415.36261(16)$ \\
        $R_{p}/R_{s}$ & $0.08004(41)$ \\
        Linear Limb Darkening, $u_{1}$ & $0.203(23)$ \\
        Eclipse 1 Midpoint [BJD$_{\text{TDB}}$] & $2458414.62237(32)$ \\
        Eclipse 2 Midpoint [BJD$_{\text{TDB}}$] & $2458416.10367(37)$ \\
        Eclipse Duration, $T_{14}$ [days] & $0.16255(47)$ \\
        Eclipse Depth, $D$ [\%] & $0.3131(62)$ \\
        $A_{1}$ [ppm] & $975(32)$ \\
        $t_{1}$ [BJD$_{\text{TDB}}$] & $2458414.5544(90)$ \\
        $A_{2}$ [ppm] & $89(22)$ \\
        $t_{2}$ [BJD$_{\text{TDB}}$] & $2458414.555(33)$ \\
        $\alpha$ [ppm] & $0(7)$ \\
        $\beta$ [ppm] & $-140(20)$ \\
        $\Pi$ [hr] & $7.56(0.13)$ \\
        Linear Ramp, $v$ [ppm/day] & $157(36)$ \\
        Linear Fit to x PRF FWHM, $PRF_{x}$ & $-0.1143(43)$ \\
        Linear Fit to y PRF FWHM, $PRF_{y}$ & $0.0451(51)$ \\
    \hline \hline
    \textbf{Derived Parameters} & \textbf{Value} \\
    \hline
    Phase Curve Amplitude, $A$ & $0.609 \pm 0.020$\\
    Phase Offset [$\degr$] & $18.7^{+2.1}_{-2.3}$\\
    Dayside Brightness Temperature, $T_{day}$ [K] & $4566^{+140}_{-136}$ \\
    Hottest Hemisphere Brightness Temperature [K] & $4636^{+145}_{-138}$ \\
    Nightside Brightness Temperature, $T_{night}$ [K] & $2556^{+101}_{-97}$ \\
    Day-Night Temperature Contrast, $A_{T}$ & $0.440^{+0.017}_{-0.016}$ \\
    \hline \hline
    \end{tabular}
    \caption{Best-fit values and errors on all free parameters in the final fit and physical parameters derived from the fit. Numbers in parentheses give $1\sigma$ uncertainties on the least significant digits. Errors on derived temperatures incorporate the error in the stellar temperatures.}
    \label{tab:params}
\end{table}

\section{Comparison to General Circulation Models}
\label{sec:gcm}
We used the GCM of \cite{Tan2019} to compare the phase curve to numerical predictions. This GCM includes the effects of cooling due to dissociation of molecular hydrogen and heating from recombination of atomic hydrogen, along with changes in the specific heat and specific gas constant due to H$_{2}$ dissociation/recombination. The dynamical core of the MITgcm solves the primitive equations of motion on a cubed-sphere grid \citep{Adcroft:2004}. We used a double-grey approximation, with one visible and one infrared band in the radiative transfer calculation \citep{Komacek:2017}, the opacity of which depends on pressure alone\footnote{The thermal opacity profile is $\log_{10}\kappa_{\rm{th}} = 0.0498(\log_{10}p)^2 - 0.1329\log_{10}p - 2.9457$ and the visible opacity profile is $\log_{10} \kappa_{\rm{v}} = 0.0478(\log_{10}p)^2 - 0.1366\log_{10}p - 3.2095$, with opacity in units of $\rm{m^2kg^{-1}}$ and pressure is in units of Pa.}. This opacity profile is the same as used in  \cite{Tan2019}. We use this simplified opacity profile for our idealized model because relevant opacities have not been calculated exactly at the temperature of KELT-9b \citep{Freedman2014} and our GCM setup is unable to fully capture dayside-to-nightside opacity differences.
We used 192 grid points in longitude and 96 in latitude, with 50 vertical levels evenly spaced in log-pressure from 1 mbar to 100 bars. We chose a model top of 1 mbar because the pressure-dependent double-grey opacity scheme used in the GCM does not apply at low pressures \citep{Rauscher_2012}. We fix the stellar $T_{\rm eff}$, $a/R_*$, and $R_p/R_*$ to the values from \citet{Gaudi2017}.  

\begin{figure*}
    \centering
    \includegraphics[width=\linewidth]{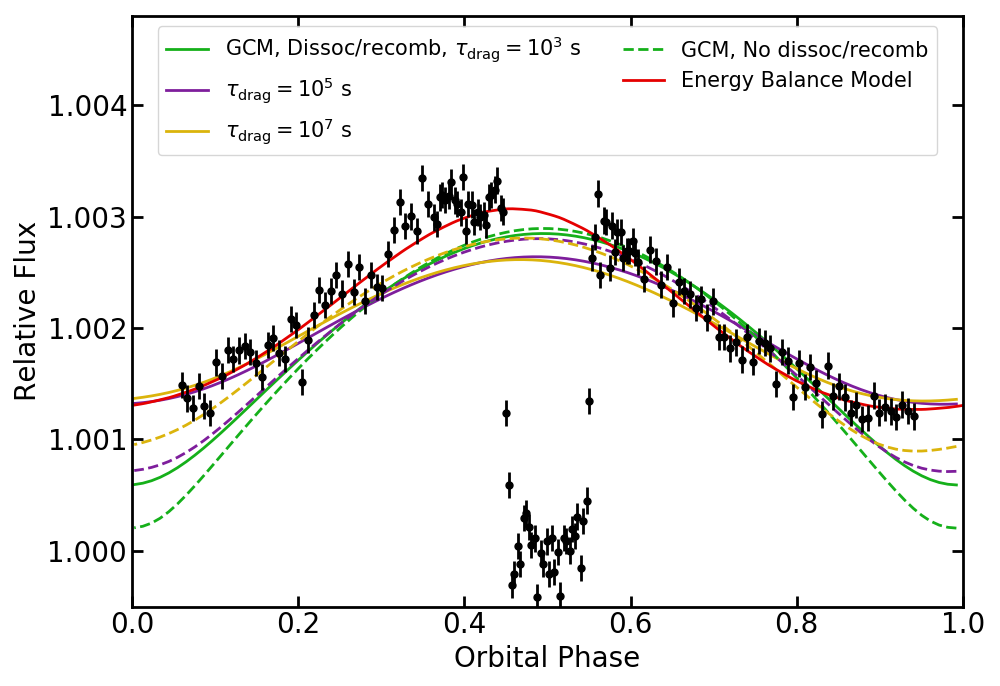}
    \caption{Phase-folded phase curve of KELT-9b (black data points). The transit at phases of 0 and 1 is omitted to better show the phase variation. Green, purple, and gold lines show GCMs with drag timescales of $10^{3}$, $10^{5}$, and $10^{7}$ seconds, respectively (Section~\ref{sec:gcm}). Solid and dashed lines indicate GCMs with and without the effects of H$_{2}$ dissociation and recombination, respectively. The red line shows the EBM including the effects of H$_{2}$ dissociation and recombination (Section~\ref{sec:ebm}).}
    \label{fig:phasegcm}
\end{figure*}

We performed multiple GCM experiments with varying frictional drag to crudely represent magnetic effects \citep{Perna_2010_1,Rauscher_2013,Rogers:2014} and/or large-scale turbulence \citep{Youdin_2010,Fromang:2016}. We used a Rayleigh drag that is linear in wind speed, $f_\mathrm{drag} = -{\bf u}/\tau_\mathrm{drag}$, where ${\bf u}$ is velocity and $\tau_\mathrm{drag}$ is the frictional drag timescale. We considered a broad range of frictional drag timescales $\tau_\mathrm{drag} = 10^3, 10^5, 10^7 \ \mathrm{s}$, to represent the unknown dipolar magnetic field strength \citep{Yadav:2017} and/or length-scale of instabilities \citep{Koll2018}. Frictional drag begins to strongly affect the circulation for $\tau_\mathrm{drag} \lesssim 10^5 \ \mathrm{s}$ \citep{Komacek:2017}, while $\tau_\mathrm{drag} = 10^7 \ \mathrm{s}$ represents very weak drag. For each assumed drag timescale, we ran GCM experiments both including and not including the effects of H$_{2}$ dissociation, resulting in six separate GCM experiments. Our simulations with weak drag have an eastward equatorial jet, while our simulations with strong drag have day-to-night flow at photospheric levels. We compare the simulated phase curves to the observations in Figure~\ref{fig:phasegcm}. 

We compare our observations to the models using the derived amplitude listed in Table~\ref{tab:params}. The observed low amplitude $A=0.609 \pm 0.020$ 
indicates significant heat redistribution from the hot dayside. Overall, we find that simulations including the impact of H$_{2}$ dissociation/recombination and with relatively weak drag provide a better match to the phase curve amplitude, while those without H$_{2}$ dissociation/recombination and/or with strong drag predict too-large amplitudes and too-cold nightsides.

\begin{figure}
    \centering
    \includegraphics[width=\linewidth]{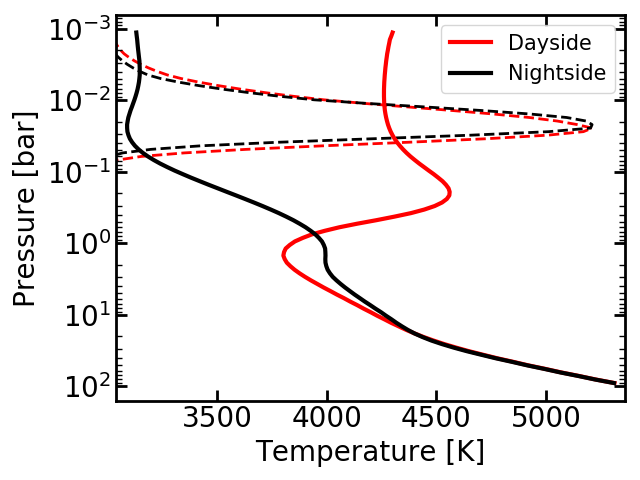}
    \caption{
    Temperature-pressure profiles from the GCM run with $\tau_{\text{drag}}=10^{7}$ and including the effects of H$_{2}$ dissociation. The dayside and nightside averaged profiles are shown in red and black, respectively. Dashed lines show contribution functions for the center of the \textit{Spitzer} 4.5~$\mu$m bandpass from 1D modeling using HELIOS \citep{malik19}. The photosphere is at $\sim10$~mbar on both the day- and nightsides.
    }
    \label{fig:matejmods}
\end{figure}

Recent work has suggested that, in many cases, differences in opacity on the day- and nightsides of hot Jupiters may lead to different pressures being probed through the phase curve in the 4.5~$\mu$m bandpass \citep{Dobbs2017}. This can complicate an otherwise straightforward determination of the amount of heat transport in the atmosphere, because the observed day-night temperature contrast may be partially due to the changing photospheric pressure. To determine the impact this could have on our measurements, we modeled the dayside and nightside emission using the 1D radiative transfer code HELIOS \citep{malik19}. We used dayside and nightside temperature-pressure (T-P) profiles from the GCM run with $\tau_{\text{drag}}=10^{7}$ and including the effects of H$_{2}$ dissociation/recombination. Figure~\ref{fig:matejmods} shows the 
contribution functions for the \textit{Spitzer} bandpass using these T-P profiles. We found that the 4.5~$\mu$m photosphere was at a pressure of $\approx 10$~mbar on both the dayside and the nightside. Since the dayside and nightside 4.5~$\mu$m photospheres are at approximately the same pressure, the temperature difference we observe is primarily due to horizontal heat transport.

We also observe a large phase offset of $18.7^{+2.1}_{-2.3}\degr$
. While the GCM experiments including H$_{2}$ dissociation and recombination are able to explain the small amplitude of the phase curve, none of the simulations predict the large offset we observe. The simulations predict an offset of no more than $5\degr$, which is inconsistent with our observations at $>5\sigma$ confidence.

\section{Comparison to Energy Balance Models}\label{sec:ebm}

As a second test of the impact of H$_{2}$ dissociation and recombination on the phase curve of KELT-9b, we compare our findings to the open source \texttt{Bell\textunderscore EBM}\footnote{\url{https://github.com/taylorbell57/Bell_EBM}} energy balance model \citep[EBM,][]{Bell2018}. We use this analytic model in addition to the GCM because it allows us to perform a fit to the data and retrieve parameters that can be compared for models with and without H$_{2}$ dissociation/recombination. The EBM was fit to the 
phase curve using the MCMC package \texttt{emcee} \citep{ForemanMackey2015}. In order to allow convergence in a reasonable time frame, we fixed the 4.5~$\mu$m reference pressure to $P_{0}=0.1$~bar and fixed the stellar $T_{\rm eff}$, $a/R_*$, and $R_p/R_*$ to the values from \citet{Gaudi2017}. We use $P_{0}=0.1$~bar because it is the approximate depth at which heat is deposited and re-radiated and because longer convergence times at lower pressures mean that it is unfeasible to run even a simplified EBM fit at lower pressures. We fit for the wind speed in the planet's rotating reference frame ($v_{wind}$) and the planet's Bond albedo ($A_{B}$). 
To convert the planet's temperature map into a light curve, we used a 4.5~$\mu$m stellar brightness temperature of 8287~K found using a PHOENIX stellar model with $T_{\rm eff}=10200$~K \citep{Husser2013}.

Our initial fits showed that the EBM was generally able to recover the phase offset and amplitude of the phase curve, but the fitted phase curve was too sharply peaked which resulted in an overall poor fit. To improve the fit, we considered another model including a deep redistribution term that redistributes some fraction of the absorbed stellar flux uniformly around the planet. This term mimics the deeper layers (below $\sim$10 bars) of GCMs which are nearly longitudinally isothermal as the radiative timescale increases rapidly with depth \citep[e.g.,][]{Showman2009, Rauscher_2012}. This parameter allowed the EBM to fit the data well with a reduced chi-squared of 1.4 for a model with $\approx21$~\% of the absorbed flux redistributed uniformly. The best-fit EBM is shown in Figure~\ref{fig:phasegcm}.

The model including the effects of H$_{2}$ dissociation/recombination gives a $v_{\rm wind}$ of 6.1~km\,s$^{-1}$, which is on the same order of magnitude as expected for typical ultra-hot Jupiters and is similar to the $\approx 5$~km wind speed in our GCM \citep{Koll2018}
. Meanwhile, neglecting the effects of H$_{2}$ dissociation/recombination requires an unphysically high wind speed of 67~km\,s$^{-1}$ to explain the observed heat redistribution, which is further evidence of the impact of H$_{2}$ chemistry on the planet's circulation. The model also gives an albedo of $0.195 \pm 0.010$, which is similar to derived Bond albedos for other ultra-hot Jupiters \citep{Schwartz2017,Zhang2018,Kreidberg2018}.

\section{Discussion}
\label{sec:discuss}

The most striking result from the KELT-9b phase curve is the small amplitude, which shows the influence of H$_{2}$ dissociation/recombination on this planet. Recent work accounting for H$_{2}$ dissociation/recombination has demonstrated that the cooling and heating from these processes can transport significant heat, leading to reduced phase curve amplitudes on the hottest ultra-hot Jupiters \citep{Bell2018,Komacek2018}. When H$_{2}$ dissociation is not taken into account, hotter planets are expected to have less heat transport because of their shorter radiative timescales \citep[e.g.,][]{Showman2002,Cowan2011}. 
Assuming a solar composition gas and using our model photospheric pressure of $P\approx 10$~mbar, we estimate that 
KELT-9b has an extremely short radiative timescale of $\tau_{rad} \approx 30$~s \citep{Showman2002}. With that short radiative timescale, and ignoring the effects of H$_{2}$ dissociation/recombination and frictional drag, the theory of \cite{Komacek2016} and \cite{Zhang:2016} predicts a normalized dayside-to-nightside temperature contrast of $0.999$, much greater than the observed value of $0.440^{+0.017}_{-0.016}$
. Note that including the effects of frictional drag would only act to increase the dayside-to-nightside temperature contrast \citep{Komacek:2017}.

This result extends the interpretation of the phase curves of WASP-33b and WASP-103b, two ultra-hot Jupiters which were previously shown to have warm nightsides \citep{Zhang2018,Kreidberg2018}. These two planets, which both have dayside brightness temperatures around 3000~K, were hypothesized to be impacted by H$_{2}$ dissociation/recombination \citep{Bell2018,Komacek2018}. The extreme irradiation of KELT-9b enhances the impact of H$_{2}$ dissociation on the phase curve and provides stronger evidence for this process on ultra-hot Jupiters.

The reduced phase curve amplitude is well fit by both GCMs and the analytic EBM when the effects of H$_{2}$ dissociation/recombination are included. We find that relatively weak $\tau_\mathrm{drag} \ge 10^5 \ \mathrm{s}$ is required to match the nightside flux, but strong drag better explains the hot dayside. Additionally, none of the GCMs reproduce the large offset we observe. The large offset could be due to MHD effects that are not currently accounted for in the GCM used in this work. Future work investigating how magnetic effects influence both the phase curve offset and amplitude \citep[e.g.,][]{Rogers:2014,rogers17,Hindle:2019aa} could shed light on the remaining discrepancies between the Spitzer observations and GCMs.

\acknowledgements
This work is based on observations made with the \textit{Spitzer Space Telescope}, which is operated by the Jet Propulsion Laboratory, California Institute of Technology under a contract with NASA. Support for this work was provided by NASA through an award issued by JPL/Caltech. M. Mansfield acknowledges funding from a NASA FINESST grant. T.D.K. and I.W. acknowledge funding from the 51 Pegasi b Fellowship in Planetary Astronomy sponsored by the Heising-Simons Foundation. X.T. acknowledges support from the European community through the ERC advanced grant EXOCONDENSE (PI: R.T. Pierrehumbert). J.M.D. acknowledges that the research leading to these results has received funding from the European Research Council (ERC) under the European Union's Horizon 2020 research and innovation program (grant agreement no. 679633; Exo-Atmos). J.M.D acknowledges support by the Amsterdam Academic Alliance (AAA) Program. This work was completed with resources provided by the University of Chicago Research Computing Center. Part of the research was carried out at the Jet Propulsion Laboratory, California Institute of Technology, under contract with the National Aeronautics and Space Administration.

%\bibliography{main.bib}

\end{document}